\documentclass[twocolumn]{aastex701}
\usepackage{graphicx}
\usepackage{amsmath}
\usepackage{amssymb}
\usepackage{hyperref}
\usepackage{subfigure}
\usepackage{cases}
\usepackage{mathrsfs}
\usepackage{threeparttablex}
\usepackage{longtable}
\usepackage{amssymb}
\usepackage{pifont}
\usepackage{epstopdf}
\usepackage{xcolor}
\usepackage[all]{hypcap}
\usepackage{mwe,tikz}
\usepackage{bm}
\usepackage{multirow}
\usepackage{xspace}
\usepackage{rotating}
\usepackage{CJK}
\usepackage{url}
\usepackage{upgreek}
\usepackage{booktabs}
\usepackage{textcomp}
\usepackage{makecell}
\usepackage{enumitem}
\usepackage{csquotes}
\usepackage{tabularx}

\newcommand{\e}{\text{e}}

\newcommand{\p}{\partial}

\begin{document}

\title{Discovery of a Short-Duration Fast X-ray Transient Associated with a Short Gamma-Ray Burst}

\author[orcid=0000-0002-5596-5059]{Yi-Han Iris Yin$^{*}$}
\affiliation{Hong Kong Institute for Astronomy and Astrophysics, University of Hong Kong, Pokfulam Road, Hong Kong, China}
\affiliation{Department of Physics, University of Hong Kong, Pokfulam Road, Hong Kong, China}
\email[show]{iris.yh.yin@connect.hku.hk}
\thanks{These authors contributed equally to this work}

\author[orcid=0000-0001-6223-840X]{Run-Duo Liang$^{*}$}
\affiliation{National Astronomical Observatories, Chinese Academy of Sciences, Beijing 100101, China}
\affiliation{University of Chinese Academy of Sciences, Chinese Academy of Sciences, Beijing 100049, China}
\email[hide]{liangrd@bao.ac.cn}

\author[0000-0002-9615-1481]{Hui Sun}
\affiliation{National Astronomical Observatories, Chinese Academy of Sciences, Beijing 100101, China}
\email[hide]{hsun@bao.ac.cn}

\author[0000-0002-5485-5042]{Jun Yang}
\affiliation{Institute for Astrophysics, School of Physics, Zhengzhou University, Zhengzhou 450001, China}
\email[hide]{yangjun@zzu.edu.cn}

\author[0000-0001-5249-2797]{Haiwu Pan}
\affiliation{National Astronomical Observatories, Chinese Academy of Sciences, Beijing 100101, China}
\email[hide]{panhaiwu@nao.cas.cn}

\author[orcid=0000-0001-7199-2906]{Yong-Feng Huang}
\affiliation{School of Astronomy and Space Science, Nanjing University, Nanjing 210023,  China}
\affiliation{Key Laboratory of Modern Astronomy and Astrophysics (Nanjing University), Ministry of Education, China}
\email[hide]{hyf@nju.edu.cn}

\author[0000-0002-7077-7195]{Jirong Mao} 
\affiliation{Yunnan Observatories, Chinese Academy of Sciences, 650011 Kunming, Yunnan Province, China} 
\email[hide]{jirongmao@mail.ynao.ac.cn} 

\author[0000-0003-4111-5958]{Bin-Bin Zhang}
\affiliation{School of Astronomy and Space Science, Nanjing University, Nanjing 210023, China}
\affiliation{Key Laboratory of Modern Astronomy and Astrophysics (Nanjing University), Ministry of Education, China}
\email[hide]{bbzhang@nju.edu.cn}

\author[0000-0002-0170-0741]{Cui-Yuan Dai}
\affiliation{School of Astronomy and Space Science, Nanjing University, Nanjing 210023, China}
\affiliation{Key Laboratory of Modern Astronomy and Astrophysics (Nanjing University), Ministry of Education, China}
\email[hide]{cydai@smail.nju.edu.cn}

\author[0000-0002-5881-335X]{Xiang-Yu Wang}
\affiliation{School of Astronomy and Space Science, Nanjing University, Nanjing 210023, China}
\affiliation{Key Laboratory of Modern Astronomy and Astrophysics (Nanjing University), Ministry of Education, China}
\email[hide]{xywang@nju.edu.cn}

\author[0000-0001-8266-3024]{Weimin Yuan}
\affiliation{National Astronomical Observatories, Chinese Academy of Sciences, Beijing 100101, China}
\email[show]{wmy@nao.cas.cn}

\author[orcid=0000-0002-9725-2524]{Bing Zhang}
\affiliation{Hong Kong Institute for Astronomy and Astrophysics, University of Hong Kong, Pokfulam Road, Hong Kong, China}
\affiliation{Department of Physics, University of Hong Kong, Pokfulam Road, Hong Kong, China}
\email[hide]{bzhang1@hku.hk} 

\begin{abstract}

Extragalactic fast X-ray transients (EFXTs) are an emerging population of high-energy transients with diverse physical origins. Only a few observed EFXTs are associated with compact object mergers, and the known events generally display X-ray emission lasting much longer than their associated short gamma-ray bursts (GRBs). Here we report EP260411a, an $\sim8$-s EFXT discovered in archival Einstein Probe Wide-field X-ray Telescope (EP-WXT) observations through a time-targeted search for X-ray counterparts of short GRBs. This fast X-ray transient is temporally coincident with short GRB 260411A, with both the pulse width and peak time showing clear anticorrelations with the characteristic photon energy across the gamma-ray and soft X-ray bands. Joint EP-WXT and Fermi-GBM spectral fitting further shows that the broadband emission can be described by a cutoff power-law model with an additional low-energy spectral break, suggesting a common origin for the single-pulse X-ray and gamma-ray emission. The resulting properties place the event within the Type I GRB region in the $E_{\rm p,z}$-$E_{\rm iso}$ plane, supporting its classification as a compact object merger event. Unlike previously known merger-driven Type I EFXTs with long-lasting X-ray emission, EP260411a reveals that the Type I EFXT population can also extend to the short timescale of a few seconds. Its combination of short duration and hard spectrum occupies a sparsely explored region of the EFXT parameter space, providing a new observational link between short-duration EFXTs and compact object mergers.

\end{abstract}

\section{Introduction} 
\label{sec:intro}

Extragalactic fast X-ray transients (EFXTs) are cosmological events whose emission is predominantly detected in the X-ray band, with durations ranging from milliseconds to hours, whose physical origins are diverse \citep[][B. Zhang et al., 2026, in preparation]{Soderberg2008,Jonker2013,Glennie2015,Irwin2016,Bauer2017,Lin2018,Xue2019,Alp2020,Lin2022,QV2022,Quirola2024A&A,Sun2025NSR}. A rapidly growing population of EFXTs has been discovered by the Wide-field X-ray Telescope (WXT) onboard the Einstein Probe (EP), thanks to its large field of view and unprecedented sensitivity in the soft X-ray band \citep{Yuan2022, Yuan2025}, offering new insights into their observational properties and physical origins. A majority of EFXTs detected by EP exhibit X-ray luminosities of $\sim10^{46}$–$10^{50}~{\rm erg~s^{-1}}$ on the minimum variability timescales of seconds to minutes, suggesting that their emission originates from relativistic outflows. 30\%--50\% of EFXTs are associated with gamma-ray bursts (GRBs; Q. Wu et al., in preparation; J. Yang et al., in preparation), indicating a close connection in the physical origins of these two high-energy transient populations \citep{OConnor2025ApJ, Guo2025ApJ}. Analogous to the physically motivated Type I/II classification established for GRBs \citep{zhangnature06,zhang09}, EFXTs may likewise be classified into, but are not limited to, Type I and Type II progenitor systems, associated with compact object mergers and the core collapse of massive stars, respectively (see B. Zhang et al., 2026, in preperation, for a detailed discussion). Indeed, several EFXTs have been found to be associated with supernovae, including EP240414a \citep{Sun2025, Zheng2025, Bright2025, vanDalen2025, Hamidani2025}, EP240506a \citep{Liang2026}, EP240801a \citep{Jiang2025}, EP250108a \citep{LiWX2025, Zhu2025MNRAS}, EP250304a \citep{Cotter2026, Song2026}, EP250827b \citep{Srini2026}, and EP260321a \citep{Yuan2026, OConnor2026, Rastinejad2026, Antonio2026, Chen2026}, providing direct evidence for a Type II origin for the majority of the EFXT population.

On the other hand, even though Type I EFXTs have been predicted long ago \citep{Zhang2013ApJ}, they are rarely identified among the EFXT population. 
Previous X-ray monitors have jointly detected short-duration GRBs and X-ray transients with properties resembling those expected for Type I EFXTs, yet their physical origins often remained uncertain. The Wide Field Camera (WFC) instrument onboard the BeppoSAX X-ray observatory \citep{Jager1997} detected a few EFXTs lasting less than 10 s in the 2--30 keV band \citep{BeppoSAX2000, BeppoSAX2026}, but these events were likely associated with massive star collapse \citep{GRB991014, GRB980326}. The Wide-field X-ray Monitor (WXM) instrument onboard the High Energy Transient Explorer 2 (HETE-2) satellite \citep{Lamb2004} also detected several short GRBs with X-ray emission extending down to 2 keV \citep{HETE22006}, including GRBs 010326B, 020531 \citep{GRB020531}, 050709 \citep{GRB050709}, and 060121 \citep{GRB060121}. Notably, GRBs 050709 and 060121 displayed a long, soft bump following their initial short, hard spikes, a feature also commonly observed in Type I GRBs. However, the physical origin of the extended soft emission remains inconclusive \citep{GRB050709}. LXT230307a, an EFXT detected by the EP pathfinder LEIA \citep{LEIA2022ApJ,LEIA2023RAA}, represents a firm example \citep{Sun2025NSR}. Its X-ray emission lasted approximately five times longer than the associated Type I GRB 230307A, whose merger origin was independently established by the detection of a kilonova \citep{Levan2024, Yang2024Nat}. Two other Type I EFXTs, EP250704a \citep{Li2026} and EP260527a \citep{EP260527a_GCN}, are associated with classical short GRBs but exhibit soft X-ray emission extending for several minutes. Such prolonged X-ray emission may arise from long-lasting central engine activity following the merger \citep{Li2026}. EP250207b and LXT240402a have also been proposed as Type I candidates based primarily on the properties of their apparent host galaxies \citep{Becerra26, Jonker26}, although associations with more distant host galaxies cannot yet be conclusively excluded \citep{vanhoof26}. With the currently limited sample, Type I EFXTs do not appear to occupy a distinct region of X-ray parameter space: the known candidates generally last for minutes, and share similar X-ray durations and other properties with the much more numerous Type II EFXTs. Intriguingly, however, EP250704a exhibits an initial short X-ray spike temporally and spectrally coincident with the short GRB 250704B \citep{Li2026}, indicating that an on-axis relativistic jet can produce not only a short GRB but also a short-duration fast X-ray transient as its natural extension to lower energies. Such short-duration X-ray emission may therefore provide an additional observational signature of Type I EFXTs.

\begin{figure*}
  \centering
    \includegraphics[width=0.7\linewidth]{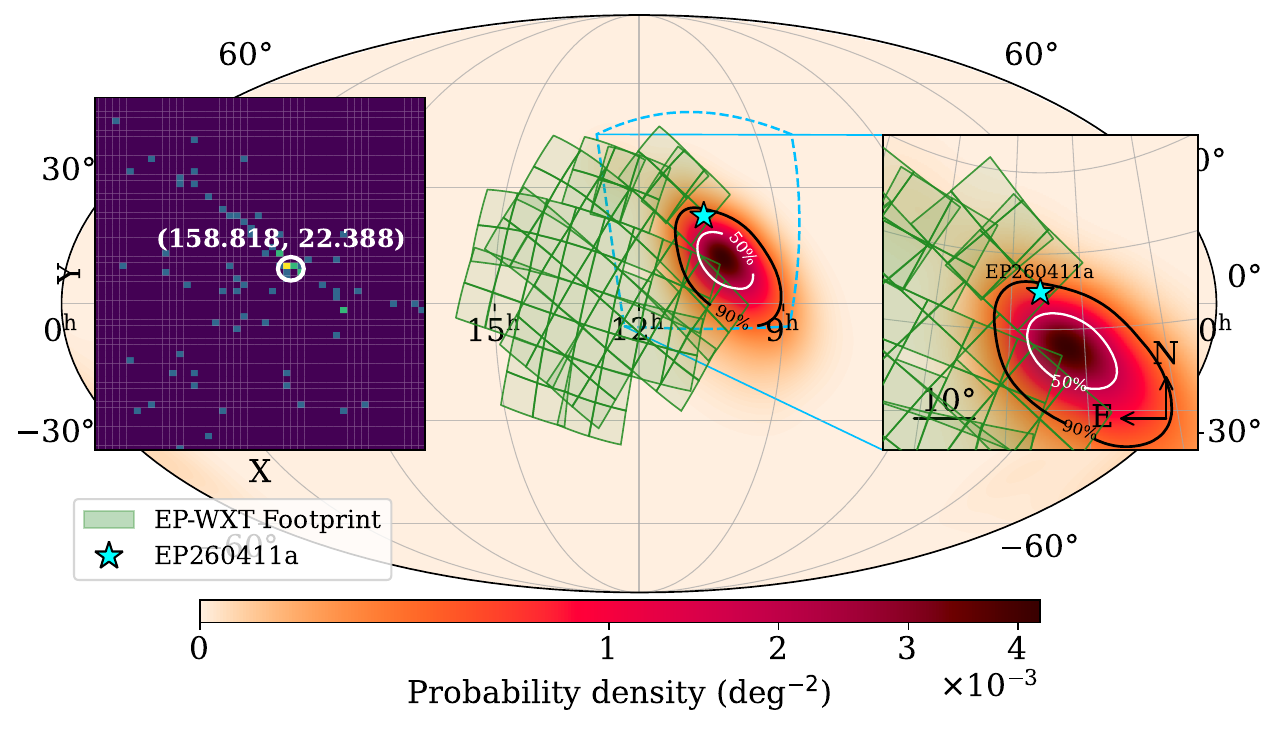}
    \includegraphics[width=0.45\linewidth]{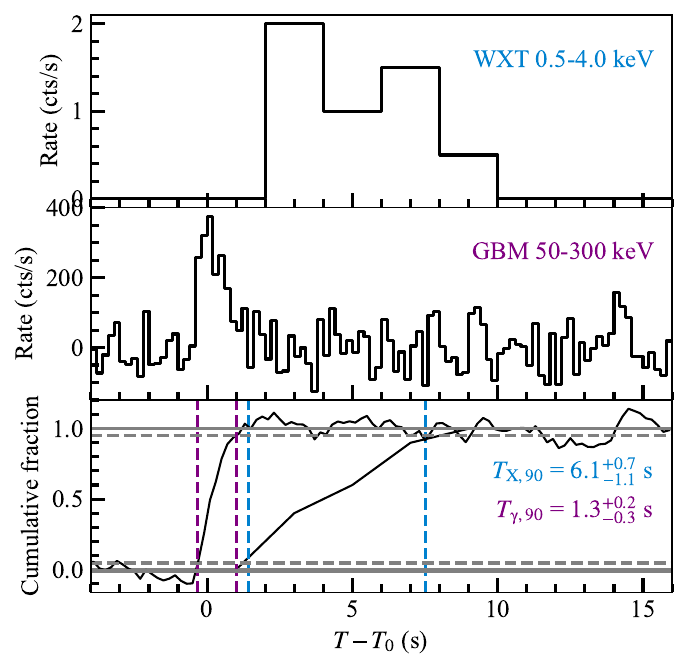}
    \includegraphics[width=0.45\linewidth]{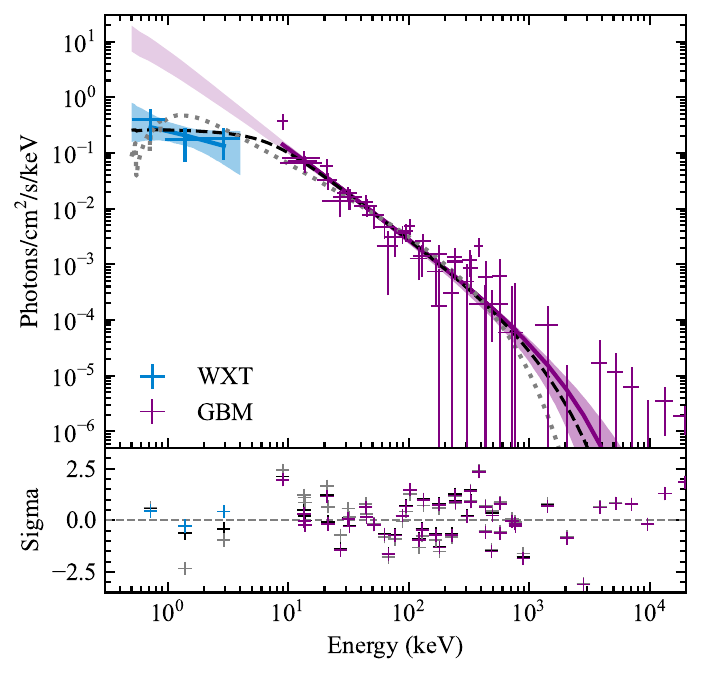}
    \caption{\textit{Upper:} Relative localization of EP260411a and GRB 260411A. The Fermi-GBM localization of GRB 260411A is shown in red, with a 90\% credible region spanning 672.06~deg$^{2}$. The EP-WXT footprints within $T_{0}+[-200,+600]$ s are shown as green rectangles, and the position of EP260411a is marked by a cyan star. EP260411a lies within the 82\% credible region of the Fermi-GBM localization. The insert shows a zoomed-in view of the 20-s stacked WXT CMOS 39 image of EP260411a. The image is binned by approximately 35 pixels, and the white circle marks the source extraction region with a radius of 67 pixels. \textit{Lower left:} EP-WXT and Fermi-GBM light curves and $T_{90}$ measurements of EP260411a/GRB 260411A. The top and middle panels show the EP-WXT 0.5--4.0 keV and Fermi-GBM 50--300 keV light curves with bin sizes of 2.0 s and 0.2 s, respectively. The bottom panel shows the normalized cumulative light curves. The gray horizontal solid (dashed) lines mark 0\%--100\% (5\%--95\%) of the total accumulated counts levels. The blue (purple) vertical dashed lines denote the $T_{\rm X,90}$ ($T_{\gamma,90}$) interval. \textit{Lower right:} Photon spectra of EP260411a/GRB 260411A obtained from independent and joint spectral fits of the EP-WXT and Fermi-GBM data. The best-fit absorbed PL, CPL, absorbed CPL, and absorbed CSBPL models are indicated by the blue solid, purple solid, gray dotted and black dashed curves, respectively. All error bars mark the 1$\sigma$ uncertainties.}
    \label{fig:detection}  
\end{figure*}

Despite theoretical expectations, short-duration EFXTs are particularly hard to detect with EP-WXT. One plausible way to identify short Type I EFXTs is through their association with short Type I GRBs, whose prompt X-ray counterparts may appear as EFXTs with durations consistent with a natural low-energy extension of the short GRB emission. However, the limited number of photons accumulated over only a few seconds, owing to the relatively small effective area of EP-WXT, makes these events difficult to be triggered by the onboard detection system. Consequently, no such short-duration Type I EFXT has been detected by EP hitherto.

In this Letter, we report the discovery of EP260411a, an 8-s EFXT associated with the short GRB 260411A, through a time-targeted search of archival EP-WXT data. The structure of this Letter is organized as follows. Section \ref{sec:discovery} describes the time-targeted search and the data reduction for EP260411a and GRB 260411A. In Section \ref{sec:EP260411a}, we present detailed association and spectral analyses of EP260411a and GRB 260411A, and compare the X-ray properties of EP260411a with those of other EFXTs. Finally, we summarize the main findings and discuss the physical interpretations of different subtypes of Type I EFXTs in Section \ref{sec:sum}.

\section{Targeted Search and Data Reduction}
\label{sec:discovery}

\subsection{Sample Selection and Time-Targeted Search}

We collected 166 short GRBs reported through the General Coordinates Network (GCN) and Fermi GBM Burst Catalog\footnote{\url{https://heasarc.gsfc.nasa.gov/w3browse/fermi/fermigbrst.html}} between 2024 January 9 and 2026 June 1 and conducted a systematic search for their X-ray counterparts in archival EP-WXT observations (Yin et al., in preparation). Among the 166 short GRBs, 33 are well localized, with localization uncertainties smaller than $10 \arcsec$, while the remaining 133 have larger error regions ranging from several square arcminutes to thousands of square degrees. Of the 33 well-localized GRBs, three have WXT coverage within $T_0-200$ to $T_0+600$ s, including the previously identified EP240807a, EP250704a and EP260527a. The detected X-ray counterparts, however, all exhibit extended emission lasting at least an order of magnitude longer than their associated short GRBs. For the 133 GRBs with larger localization uncertainties, we first performed an image-based search within the overlap between the WXT field of view and the GRB localization region over $T_0-200$ to $T_0+600$ s. Among candidates with an image significance above 5$\sigma$, none could be confidently confirmed as a real transient or exhibited a clear temporal structure in its light curve. This search can be insensitive to short-duration X-ray counterparts, whose signals may be submerged when integrated over a much longer imaging interval.

To find faint, short-duration transients with significance near the detection threshold, we further performed a time-targeted search around the known trigger times ($T_0$) of the 133 GRBs with large localization uncertainties. For each burst, we applied a 20-s interval centered on $T_0$ and selected the WXT CMOS detectors that continuously covered the GRB localization region throughout the entire interval, yielding 6 GRBs with WXT coverage. We then extracted a stacked image over this time interval for each WXT CMOS detector and searched for uncataloged point sources. For each image source candidate, we also examined the single-frame CMOS light curves to identify and exclude source candidates exhibiting significant single-frame spikes, which are very likely caused by cosmic-ray events. After careful manual validation of the point source candidates, EP260411a was the only remaining candidate consistent with an astrophysical transient, exhibiting a point-source-like excess in the image and a structured light curve around the trigger time of the short GRB 260411A, as shown in Figure \ref{fig:detection}.

\begin{figure*}
  \centering
    \includegraphics[width=0.46\linewidth]{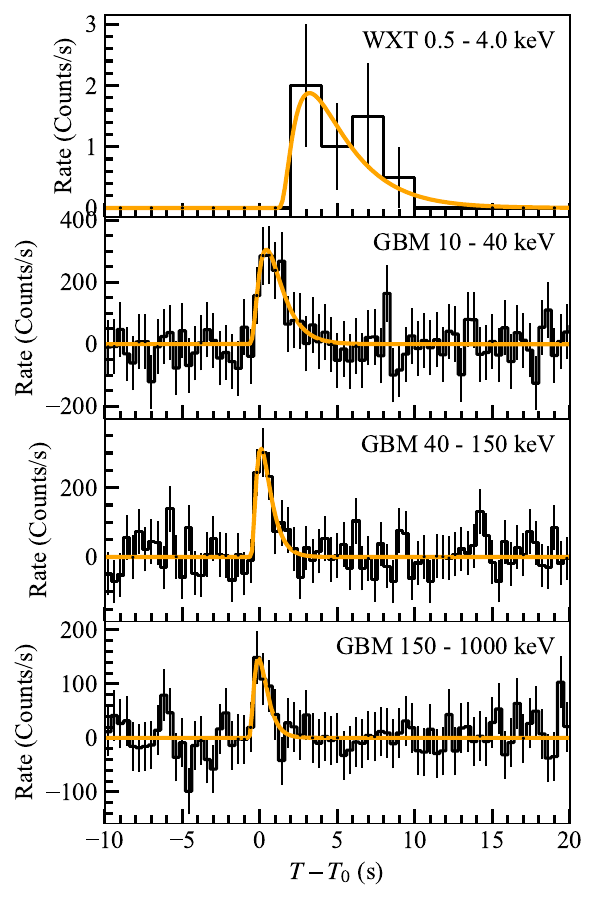}
    \includegraphics[width=0.45\linewidth]{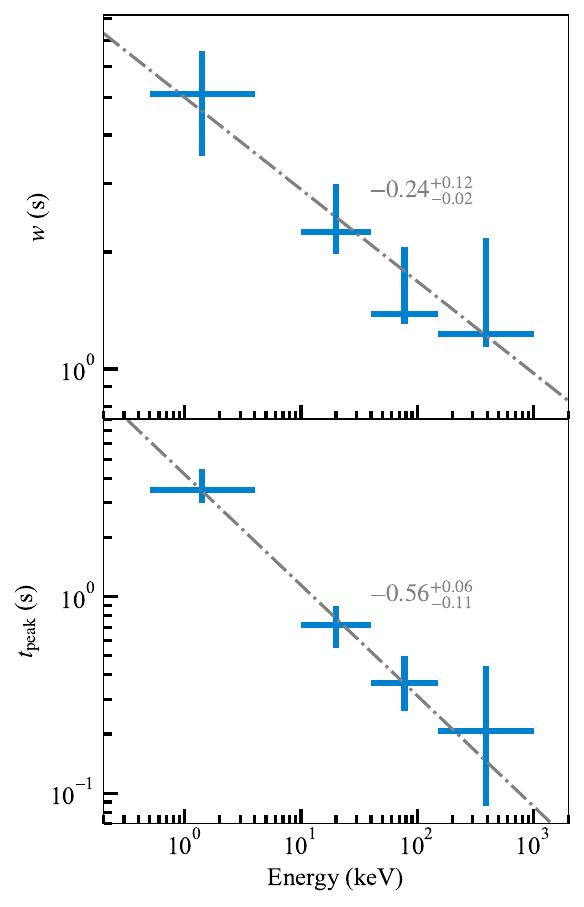}
    \caption{Multiwavelength light curves of EP260411a/GRB 260411A and the energy dependence of the pulse width and peak time. The WXT light curve is binned at 2.0 s, and the GBM light curves at 0.4 s. The yellow curves represent the best-fit FRED shape models. The gray dashed-dotted lines represent best-fit PL relations of the pulse width and peak time as functions of energy. All error bars mark the 1$\sigma$ confidence level. }
    \label{fig:consistency}  
\end{figure*}

\subsection{Data Reduction of EP260411a and GRB 260411A}

The source position of EP260411a is R.A. = 158.$^{\circ}$818 and decl. = 22.$^{\circ}$388 (J2000), with an uncertainty of 3\arcmin, which lies within the 82\% credible region of the Fermi-GBM localization (Figure \ref{fig:detection}). We extracted the source and background photons from a circular region with a radius of 67 pixels and an annulus region with inner and outer radii of 134 and 268 pixels, respectively (1 pixel $\sim$ 0.136\arcmin), centered on the source position. This yielded a total of 10 photons in the source region and 8 photons in the background region, corresponding to a Li-Ma significance of $S=5.26$ \citep{LiMa1983}. We then followed the standard data reduction procedures implemented in the WXT Data Analysis Software (\texttt{WXTDAS}; Y. Liu et al., in preparation) and the calibration database \citep[CALDB;][]{Cheng2025} to generate cleaned event files and instrumental response products.

The short GRB 260411A was detected by the Fermi Gamma-ray Burst Monitor \citep[Fermi-GBM;][]{Meegan2009} at $T_0=$ 2026-04-11T06:36:28, with a calculated location of RA=153.6, Dec=12.0 (J2000) and uncertainty of around 9.6 degrees \citep{2026GCN.44261....1F}. We retrieved the time-tagged event data and selected the sodium iodide (NaI) detectors n0, n1, n9, which have the smallest viewing angles relative to the GRB source direction among the 12 NaI detectors, together with b0, the brightest bismuth germanium oxide detector. Data reduction and analysis were performed using \texttt{heapy} \citep{heapy_cite}, following the standard procedures described in \citet{Yang2022} and \citet{Yin2024}.

The lower left panel of Figure \ref{fig:detection} presents the extracted net X-ray light curve of EP260411a with a $T_{\rm X,90}=6.1_{-1.1}^{+0.7}$ s and the gamma-ray light curve of GRB 260411A with a $T_{\rm \gamma,90}=1.3_{-0.3}^{+0.2}$ s. Both light curves exhibit a fast-rise and exponential-decay (FRED) profile.

\section{Data Analysis}
\label{sec:EP260411a}

\subsection{Association Assessment}

To assess the likelihood of a chance association between EP260411a and GRB 260411A, we estimate the probabilities of temporal and spatial coincidence under the null hypothesis that the two events are unrelated. Since the EP-WXT search was performed following the detection of GRB 260411A by Fermi-GBM, we condition our analysis on the GBM trigger and evaluate the probability of detecting an unrelated EFXT candidate coincident with the GRB in time and sky position with EP-WXT. We obtain an upper limit on the chance-coincidence probability of $P_{\rm chance}<2.73\times10^{-3}$, which corresponds to $P_{\rm chance,global}<1.62\times10^{-2}$ after accounting for a trial factor of six. The detailed calculations are presented in Appendix \ref{sec:Pcc}.

We further examined the temporal consistency between EP260411a and GRB 260411A. As shown in Figure \ref{fig:consistency}, the light curves in different energy bands can each be well described by a single FRED pulse. We fitted the X-ray and gamma-ray light curves, binned at 2.0 s and 0.4 s, respectively, using the empirical pulse model of \citet{Norris2005}:
\begin{equation}
I(t)=N/\exp\left[\frac{\tau_r}{t-t_s}+\frac{t-t_s}{\tau_d}\right],
\end{equation}
where $N$ is the normalization, $t_s$ is the pulse onset time, and $\tau_r$ and $\tau_d$ characterize the rise and decay of the pulse, respectively. Defining $\mu=\sqrt{\tau_r/\tau_d}$, the pulse reaches its maximum at $t_{\rm peak}=t_s+\sqrt{\tau_r\tau_d}$, and its width between the two $1/e$ intensity points is $w=\tau_d\sqrt{1+4\mu}$.
The 1$\sigma$ uncertainties of $t_{\rm peak}$ and $w$ were estimated through Monte Carlo simulations, assuming Gaussian distributions for the fitted parameters. Since the adopted trigger time $T_0$ occurs after the onset of the high-energy pulse, we shifted all measured peak times by 0.3 s to avoid negative values relative to $T_0$.

Both the pulse width and peak time exhibit clear energy dependence. Power-law fits yield $w\propto E^{{-0.24}_{-0.02}^{+0.12}}$ and $t_{\rm peak}\propto E^{{-0.56}_{-0.11}^{+0.06}}$, indicating that the pulse becomes broader and peaks progressively later toward lower energies. Similar energy-dependent pulse evolution has also been reported for GRB 060218 \citep{Liang2006}. The energy-peak time and energy-width anticorrelations support a common origin of EP260411a and GRB 260411A, with the X-ray transient representing the longer-duration, lower-energy extension of the short GRB \citep{zhang09}.

\subsection{Spectral Analysis}

The spectrum of EP260411a was fitted with an absorbed power-law (PL) model ($N(E)\propto E^{-\Gamma}$), with the Galactic hydrogen column density fixed at $N_{\rm H}=1.87\times10^{20}~\rm{cm^{-2}}$, yielding a photon index of $\Gamma_{\rm X}={0.60}_{-0.66}^{+0.79}$. We note that additional absorption intrinsic to the source or its host galaxy may be present, but the limited photon statistics (10 source photons) do not allow us to constraint the intrinsic hydrogen column density. For GRB 260411A, whose emission is primarily detected below 1000 keV, we fitted the Fermi-GBM spectrum with a cutoff power-law (CPL) model and obtained a photon index of $\Gamma_{\gamma}=1.61_{-0.12}^{+0.12}$ and a peak energy of $E_{\rm p}=568_{-293}^{+514}$ keV. The spectrum can also be adequately described by a single power-law model, indicating that the spectral cutoff is not very well constrained by the Fermi-GBM data alone. Owing to the limited photon statistics, the spectral parameters derived from the independent EP-WXT and Fermi-GBM fits are subject to relatively large uncertainties. In Figure 1, we present the best-fit photon spectra obtained from the independent EP-WXT and Fermi-GBM spectral fits. The EP-WXT spectrum exhibits a lower photon flux and a harder photon index compared with the direct extrapolation of the Fermi-GBM CPL model to soft X-ray energies. This discrepancy may indicate additional low-energy absorption or an intrinsic spectral break between the X-ray and gamma-ray bands. We subsequently performed joint spectral fits to the EP-WXT and Fermi-GBM data using an absorbed CPL model with the hydrogen column density left as a free parameter and an absorbed smoothly broken power-law with high-energy exponential cutoff (CSBPL) model with a fixed Galactic hydrogen column density. The best-fit parameters, fitting statistics and posterior distributions are provided in Appendix \ref{sec:specfits}. The CSBPL model provides a slightly better fit to the joint spectrum, yielding photon indices of $\Gamma_1={0.11}_{-0.61}^{+0.40}$, $\Gamma_2={1.58}_{-0.16}^{+0.16}$, and a more tightly constrained peak energy of $E_{\rm p}={386}_{-178}^{+288}$ keV. These results suggest that the single-pulse X-ray and gamma-ray emission may share a common physical origin, with a low-energy spectral break between the X-ray and gamma-ray bands and a spectral peak in the gamma-ray band \citep{Ravasio2018A&A, Ravasio2019A&A, Gompertz2023, Ravasio2026}.

\begin{figure*}
  \centering
    \includegraphics[width=0.32\linewidth]{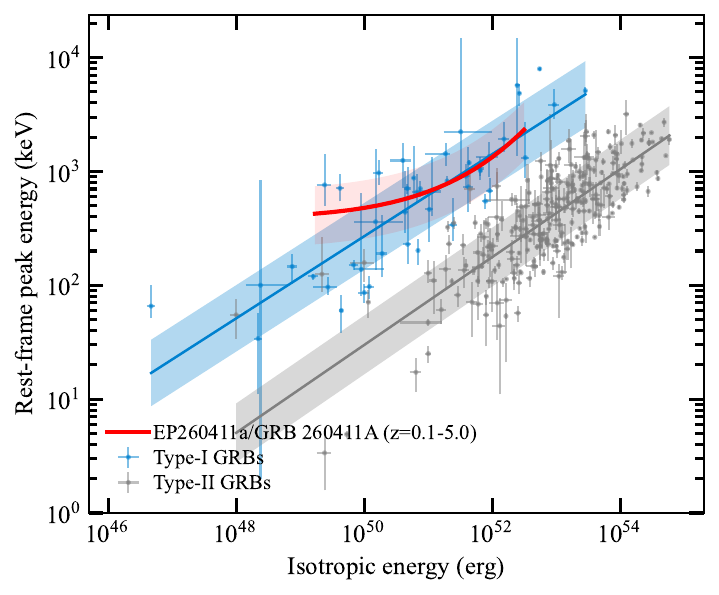}
    \includegraphics[width=0.32\linewidth]{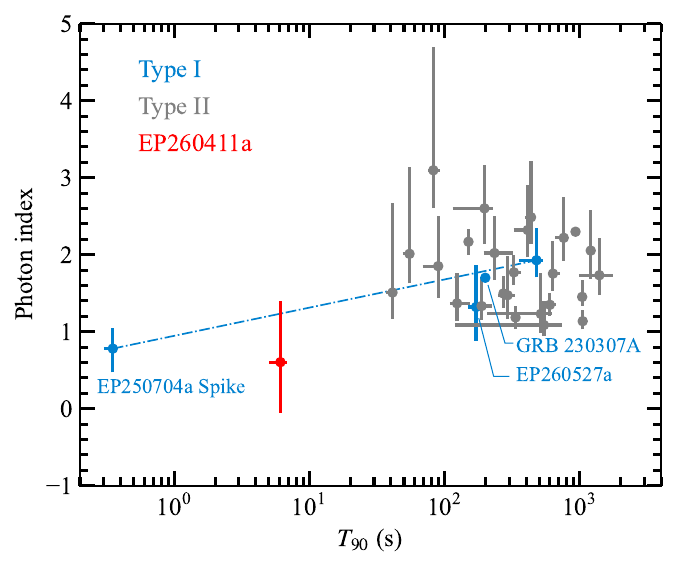}
    \includegraphics[width=0.32\linewidth]{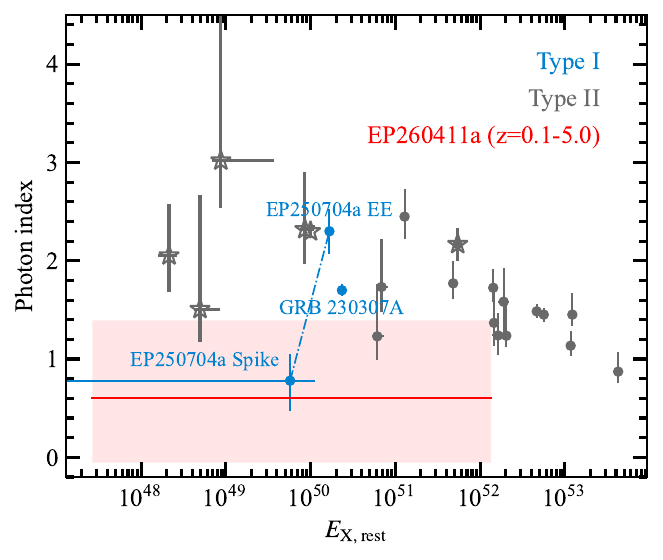}
    \caption{$E_{\rm p,z}$-$E_{\rm iso}$ plane of the GRB sample, and photon index-$T_{90}$ and photon index-$E_{\rm X,rest}$ planes of the EFXT sample. The blue and gray solid lines represent the best-fit correlations for Type I and Type II GRB populations, respectively. The red curve traces the position of GRB 260411A over the assumed redshift range, with the shaded region indicating the 1$\sigma$ uncertainty. The blue and gray data points represent the Type I and Type II EFXT populations, respectively, with stars marking Type II EFXTs associated with SNe. In the photon index-$E_{\rm X,rest}$ plane, the red shaded region indicates the 1$\sigma$ uncertainty of the photon index, while the $E_{\rm X,rest}$ range is calculated with the photon index fixed at 0.60. All error bars correspond to the 1$\sigma$ confidence level.}
    \label{fig:compare}  
\end{figure*}

To further investigate the nature of GRB 260411A, we examined its location in the $E_{\rm p,z}$-$E_{\rm iso}$ plane \citep{Amati2002}. We adopted the best-fit CSBPL parameters from the joint spectral analysis of EP260411a/GRB 260411A. Since no multiwavelength follow-up observations or redshift measurements have been reported, we calculated $E_{\rm p,z}$ and $E_{\rm iso}$ over an assumed redshift range of $z=0.1$--$5.0$. As shown in Figure \ref{fig:compare}, the track of EP260411a/GRB 260411A lies within the 1$\sigma$ region of the Type I GRB population over a broad range of redshifts, consistent with EP260411a/GRB 260411A originating from an on-axis relativistic jet launched from a compact object merger.

\subsection{Host Galaxy and Distance Constraints}

Within the $3^{\prime}$ EP localization region, we searched the REGALADE galaxy catalog~\citep{Tranin2026} and identified several potential host galaxies with redshifts spanning $z\simeq0.128$--0.366. However, given the relatively large localization area and the declining completeness beyond $z\gtrsim0.1$, neither a unique host association nor a more restrictive redshift range can be established.

We further use archival ZTF observations to constrain the distance to the burst. The localization region was observed by ZTF between $T_{0}+4.8$ and $T_{0}+73$~d. Adopting the afterglow parameters inferred for GW170817/GRB~170817A~\citep{Ryan2020} and assuming an on-axis viewing geometry, the ZTF nondetections imply $z\gtrsim0.02$. With the substantially greater depth of LSST~\citep{LSST2009,LSST2019}, a similar analysis could potentially extend this constraint to $z\gtrsim0.1$.

\subsection{EP260411a in Context}

In Figure \ref{fig:compare}, we compare the photon index, duration $T_{90}$ (middle), and rest-frame X-ray isotropic-equivalent energy $E_{\rm X,rest}$ (right) of EP260411a with those of other EFXTs. The comparison sample is adopted from Wu et al. (in preparation) and restricted to EFXTs with GRB or supernova associations. The photon index-$T_{90}$ sample includes 28 EFXTs with complete light curve detection, and the photon index-$E_{\rm X,rest}$ sample contains 23 EFXTs with redshift measurements.

In the photon index-$T_{90}$ plane, EP260411a stands out from the main distribution of Type II EFXTs, characterized by its short duration and relatively hard spectrum, similar to the initial short spike of EP250704a. The distinction of its hard spectrum from the Type II EFXT population becomes more apparent in the photon index-$E_{\rm X,rest}$ plane. EP260411a also has a relatively low X-ray isotropic-equivalent energy compared with Type II EFXTs over the assumed redshift range of $z=0.1$--$5.0$, which is naturally expected given its short duration. Such short-duration, low-flux events are particularly difficult to detect with WXT, which partly explains why this region of the EFXT parameter space has remained largely unexplored. As the first short-duration EFXT candidate discovered in association with a short GRB with soft X-ray coverage extending down 0.5 keV, EP260411a extends the observed parameter space of the EFXT population and suggests that duration, in combination with spectral hardness, may provide useful indicators for probing the physical origins of EFXTs.

\section{Summary and Discussion}
\label{sec:sum}

In this Letter, we report the discovery of EP260411a, an $\sim8$-s EFXT identified through a systematic, time-targeted search for X-ray counterparts of short GRBs in archival EP-WXT observations. EP260411a was detected around the trigger time of GRB 260411A, with a point-source detection significance of 5.26 and a light curve displaying a single FRED pulse. Both the pulse width and peak time are anticorrelated with the characteristic photon energy across the gamma-ray and X-ray bands, supporting an association between EP260411a and GRB 260411A. The X-ray spectrum of EP260411a can be fitted with an absorbed PL model with a photon index of $\Gamma_{\rm X}={0.60}_{-0.66}^{+0.79}$, while the gamma-ray spectrum of GRB 260411A exhibits a softer photon index of $\Gamma_{\gamma}=1.61_{-0.12}^{+0.12}$, suggesting a spectral break between the X-ray and gamma-ray bands. Joint spectral fitting of the X-ray and gamma-ray data with a CSBPL model yields a low-energy break at $E_{\rm b}={7}_{-3}^{+6}$ keV and a peak energy of $E_{\rm p}={386}_{-178}^{+288}$ keV.

The observed 1--$10^4$ keV fluence is $6.89_{-1.52}^{+1.51}\times10^{-7}$ $\rm{erg~cm^{-2}}$, placing this event within the Type I GRB track in the Amati relation over a broad range of assumed redshifts. EP260411a thus represents the first short-duration EFXT candidate detected by EP in association with a short GRB. Its short duration and hard spectrum distinguish it from the main distribution of Type II EFXTs in the photon index-$T_{90}$ plane and extend the observed parameter space of the EFXT population, highlighting the potential of these properties for identifying EFXTs originating from compact object mergers.

In general, Type I EFXTs could display diverse observational properties depending on their progenitor systems, jet geometries, and merger remnant types, particularly in their durations and associations with GRBs. Based on these two observational features, Type I EFXTs may be broadly divided into four subtypes\footnote{The terms ``short EFXT'' and ``long EFXT'' are used descriptively to distinguish EFXTs with relatively short and long durations, which, when associated with GRBs, correspond to a low-energy extension of the prompt pulse and extended X-ray emission, respectively. No quantitative duration boundary between the two is currently defined. Establishing such a classification will require a larger sample of EFXTs and awaits further observational criteria.}:
\begin{itemize}
\item \textit{short EFXT with GRB association.} 
Neutron star (NS)-NS mergers and NS-black hole (BH) mergers with sufficient post-merger accretion material can launch a relativistic jet and produce a short GRB when viewed on-axis. In this scenario, the short-duration EFXT originates from the same prompt jet emission as the GRB, extending the observed emission into the soft X-ray band. For the NS-NS merger, this scenario would require the remnant to collapse into a BH either promptly or after a short-lived hypermassive NS phase.
\item \textit{long EFXT with GRB association.} 
A long-duration X-ray counterpart accompanying a GRB requires additional emission beyond the initial prompt jet phase. One possible scenario involves an NS-NS merger that leaves behind a relatively long-lived remnant, such as a supramassive or stable NS \citep{Dai2006, Gao2006, Metzger2008}. NS-white dwarf (WD) mergers have also been proposed as a possible progenitor channel for long-duration merger-driven transients \citep{Yang2022, Zhang2025JHEAp}. When viewed within the jet zone, the observed X-ray emission may contain contributions from the initial jet emission and subsequent dissipation of a relativistic magnetized wind powered by the long-lived central engine.
\item \textit{short EFXT without GRB association.} 
For a merger that does not leave behind a long-lived central engine, an off-axis relativistic jet or a mildly relativistic cocoon could produce a short-duration X-ray transient without an accompanying GRB. In this scenario, the softer X-ray emission is visible over a wider angular range than the narrowly beamed gamma-ray emission, with the relativistic and geometric effects leading to observed durations ranging from a few seconds to tens of seconds \citep{Zheng2026}. Alternatively, an intrinsically weak jet might produce this observational subtype even when viewed on-axis, if its gamma-ray emission is below the detection threshold.
\item \textit{long EFXT without GRB association.} 
Although an off-axis viewing geometry can suppress the detection of the prompt GRB and lengthen the observed X-ray duration, viewing angle effects alone may not account for sufficiently long-lasting X-ray emission. In this case, a long-duration EFXT instead requires a sustained energy supply from a long-lived central engine. A natural possibility is a merger that produces a long-lived magnetar, whose relativistic magnetized wind can power the X-ray emission even after the prompt jet phase. Such a scenario was proposed by \citet{Zhang2013ApJ} and subsequently developed by \citet{Sun2017ApJ, Quirola2024A&A, chen26}. In the free zone, an observer outside the jet cone does not detect the prompt GRB but can still observe the long-lasting magnetar-wind emission. 

\end{itemize}

These subtypes qualitatively point to different physical origins discussed above, although a quantitative boundary between short- and long-duration EFXTs has yet to be established. In this classification, EP260411a provides a representative case of a short EFXT associated with a short GRB. Its association with the hard, energetic GRB 260411A favors an interpretation in which the X-ray and gamma-ray emission originate from the same on-axis relativistic jet launched following a compact object merger. Note that the longer duration observed in X-rays does not necessarily mean that the central engine duration operating in X-rays is longer than in gamma-rays, as EP260411a/GRB 260411A exhibits only a single pulse, whose duration can be determined by the emission time of the emitter rather than the central engine duration \citep{Zhang2025JHEAp}. For other cases with multiple pulses (e.g. EP250704a \citep{Li2026}), the central engine duration is indeed long. Another EFXT, EP251021b \citep{EP251021b}, may represent a similar case. Its short X-ray duration ($\sim$24 s) and a weak, untriggered GRB are consistent with a neutron star merger scenario involving the cocoon cooling emission and a structured jet viewed slightly off-axis (Dai et al., in preparation).

The short duration of EP260411a distinguishes it from previously identified merger-driven EFXTs with long-lasting X-ray emission, demonstrating that Type I EFXTs can span a wider range of timescales than currently represented by the limited sample. Such short-duration emission is consistent with a black hole as the merger remnant, formed either promptly following the merger or through the collapse of a short-lived hypermassive neutron star. As more Type I EFXTs are discovered, the short-duration, hard-spectrum region of the photon index-$T_{90}$ plane can be better explored. This will help determine whether such events occupy a region distinct from the main distribution of the current known EFXT population. A larger sample of Type I EFXTs will also allow us to establish quantitative connections between their observational properties and different progenitors, merger remnants, and viewing geometries.

\begin{acknowledgments}
We thank Qin-Yu Wu for sharing the properties of the EFXT sample. We also thank Chen-Wei Wang for checking the SVOM/GRM detection. Y.-H. I. Yin is grateful to Jin-Ping Zhu for insightful discussions on the classification of EFXT types and Run-Chao Chen for helpful suggestions.
YY and BZ acknowledge the Research Talent Hub for ITF project (RTH-ITF) from the Innovation Technology Commission of Hong Kong SAR.
YFH acknowledges the support from the Xinjiang Tianchi Program.
JM is supported by Natural Science Foundation of China 12393813, the National Key R\&D Program of China (2023YFE0101200), and the Yunnan Revitalization Talent Support Program (YunLing Scholar Project).
\end{acknowledgments}

\bibliography{ms}
\bibliographystyle{aasjournalv7}

\appendix
\restartappendixnumbering

\section{Chance-coincidence Probability for EP260411a and GRB 260411A} 
\label{sec:Pcc}

We quantify the possibility that EP260411a was detected by chance during the search for an X-ray counterpart to GRB 260411A. We calculate the temporal and spatial probabilities separately and combine them to estimate the overall chance-coincidence probability.

The temporal chance-coincidence probability $P_{\rm temporal}$ between EP260411a and GRB 260411A is taken to be the probability of detecting an unrelated short-duration X-ray transient candidate with EP-WXT within a 20-s window. To estimate this probability, we searched 1000 non-overlapping 20-s intervals in archival EP-WXT data, corresponding to a total exposure of $T=20$ ks. Applying the same detection criteria as those used for EP260411a, no candidate exhibiting a structured light curve and a detection significance exceeding that of EP260411a was identified. Assuming that unrelated transient detections are independent and randomly distributed in time at a constant average rate $r$, we adopt Poisson statistics \citep{Moroianu2023}, under which the probability of non-detection over an exposure time $T$ is $P(N=0)=e^{-rT}$. Setting $P(N=0)=0.05$, we could derive a 95\% confidence upper limit on the detection rate of $r_{95}=-\ln(0.05)/T\simeq1.5\times10^{-4}$ Hz. The probability of detecting at least one unrelated transient within a 20-s window ($\Delta t=20$ s) is constrained to $P_{\rm temporal}=1-e^{-r\Delta t}\lesssim3.0\times10^{-3}$ at the 95\% confidence level. 

We next evaluate the spatial chance-coincidence probability $P_{\rm spatial}$ based on the overlap between the EP-WXT field of view and the Fermi-GBM localization. The searchable EP-WXT area within the 90\% GBM credible region is $\sim 181.0~\rm{deg}^2$, of which $\sim 164.7~\rm{deg}^2$ lies within the 82\% credible region of the GBM localization, where EP260411a is located. Assuming that unrelated EP-WXT candidates are uniformly distributed over the searched area, we derive the spatial coincidence probability as $P_{\rm spatial}=164.7/181.0\simeq0.91$. Combining the temporal and spatial probabilities, we obtain an upper limit on the chance-coincidence probability of $P_{\rm chance}=P_{\rm temporal}P_{\rm spatial}< 2.73\times10^{-3}$, also corresponding to $P_{\rm chance,global}< 1.62\times 10^{-2}$ after accounting for a trial factor of six.

\section{Spectral Fitting Results} \label{sec:specfits}

\begin{figure*}
  \centering
    \includegraphics[width=0.46\linewidth]{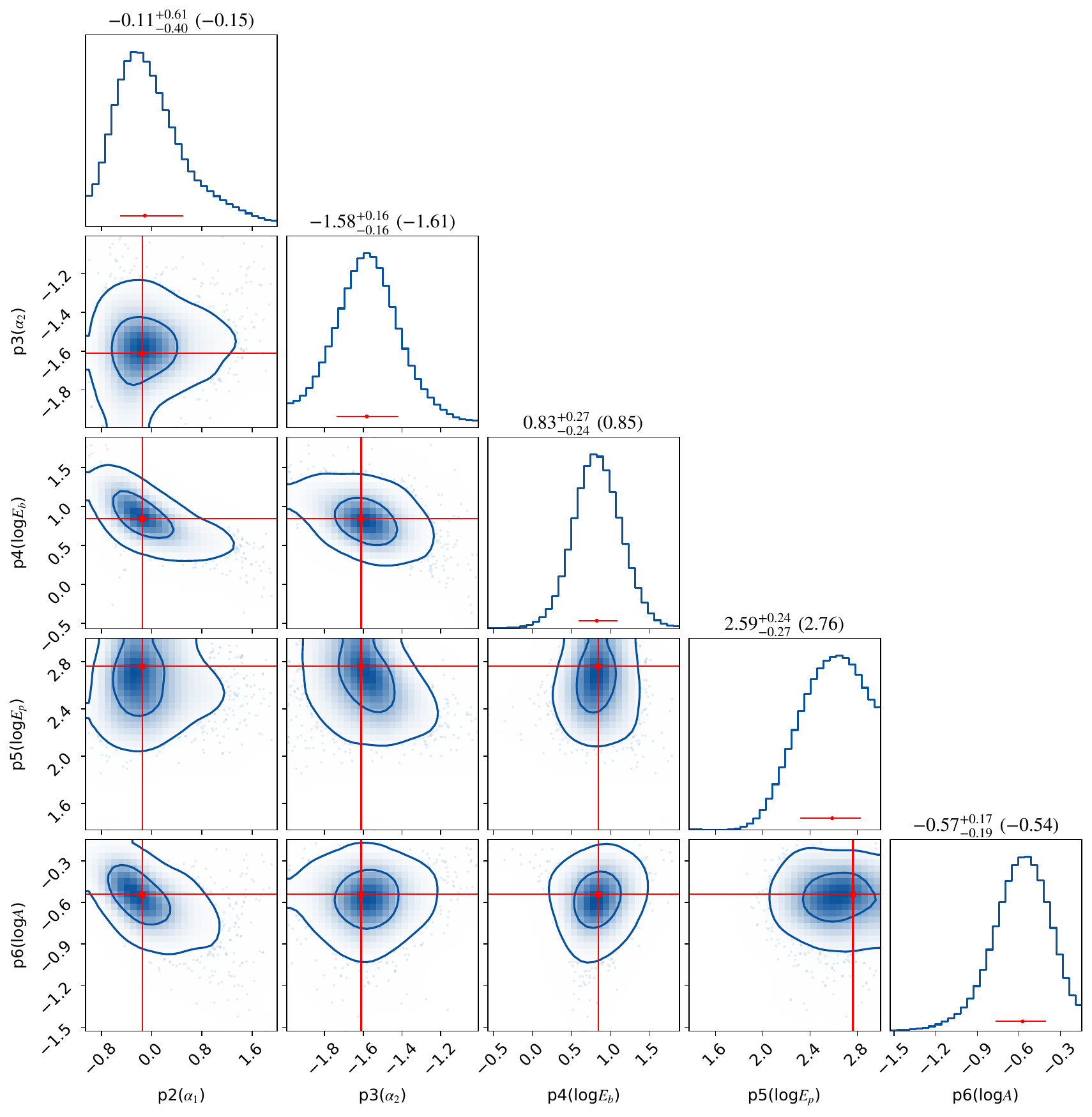}
    \includegraphics[width=0.45\linewidth]{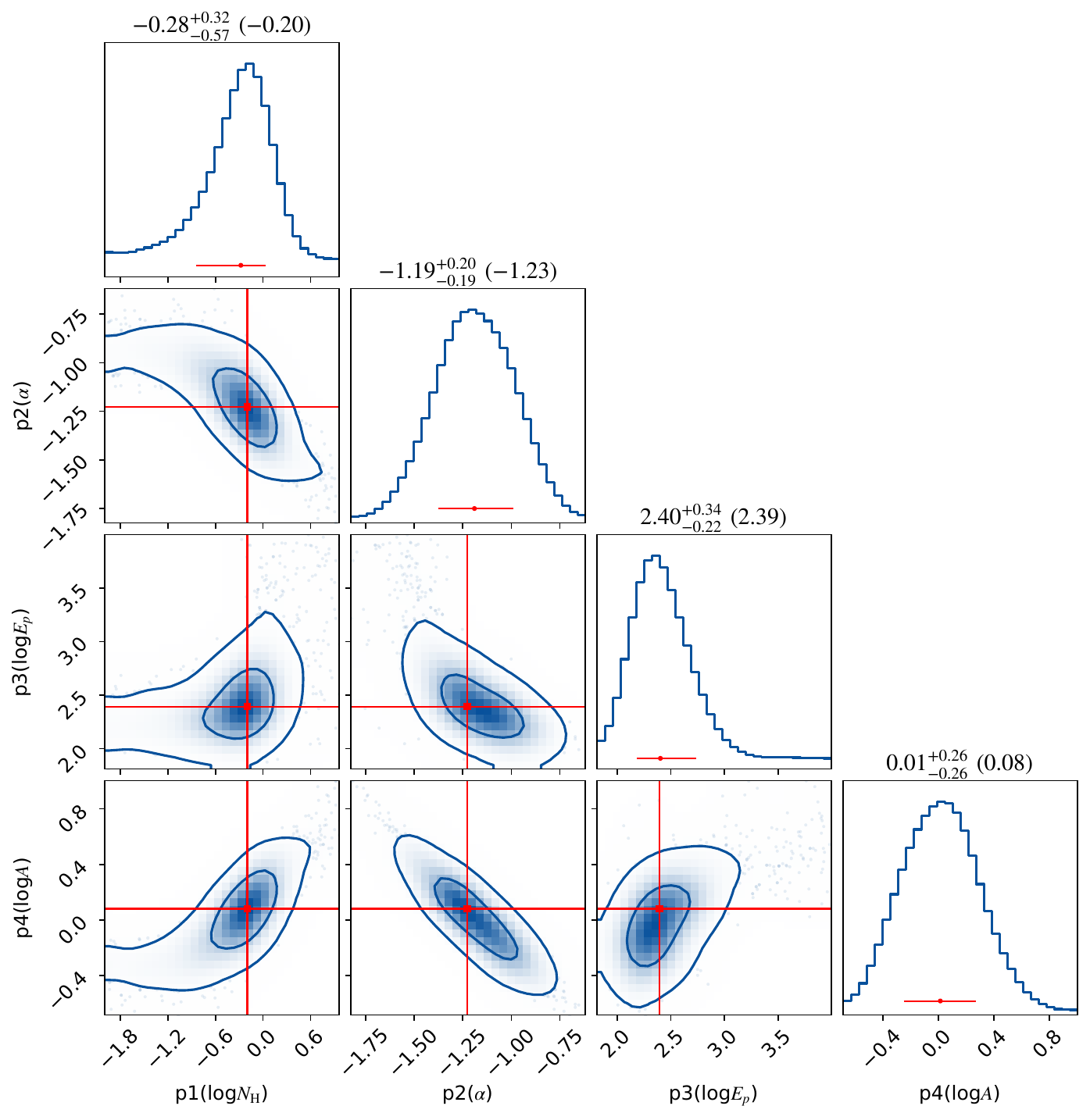}
    \caption{Corner plots of the posterior probability distributions of the parameters obtained from joint spectral fits using the \textit{tbabs*csbpl} and \textit{tbabs*cpl} models. The $N_{\rm H}$ is fixed at $1.87\times10^{20}~\rm cm^{-2}$ for the \textit{csbpl} model, while left as a free parameter for the \textit{cpl} model. The parameter $\alpha$ corresponds to the negative photon index ($\alpha=-\Gamma$). The red error bars represent 1$\sigma$ uncertainties.}
    \label{fig:corner}  
\end{figure*}

We performed time-integrated spectral fits to the EP-WXT and Fermi-GBM data over the time intervals covering the X-ray and gamma-ray pulses. For the X-ray absorption, we adopted the Tübingen-Boulder interstellar medium absorption model \citep[\textit{tbabs};][]{Wilms2000ApJ}, parameterized by the hydrogen column density $N_{\rm H}$. The spectral models considered include a power law \textit{pl}, a cutoff power law \textit{cpl}, and a smoothly broken power law with high-energy exponential cutoff \textit{csbpl}, as described in \citet{Li2026}. Spectral fitting was performed using the Python package \texttt{Bayspec} \citep{bayspec}, following the procedure described in \citet{Yin2025}. We evaluated the goodness of fit by examining the reduced statistic STAT/dof, where STAT denotes the fitting statistic and dof represents the number of degrees of freedom. Specifically, we adopted CSTAT \citep{Cash1979ApJ} for the EP-WXT spectra and PGSTAT \citep{Arnaud1996} for the Fermi-GBM spectra. Model comparisons were performed using the Bayesian information criterion \citep[BIC;][]{Schwarz1978}. The best-fit spectral parameters and corresponding fitting statistics are summarized in Table \ref{tab:spec}, and the resulting photon spectra are shown in Figure \ref{fig:detection}. The posterior probability distributions of the joint fit parameters are presented in the corner plot in Figure \ref{fig:corner}.

\begin{table*}
\centering
\scriptsize
\caption{Spectral fitting results and corresponding fitting statistics of EP260411a and GRB 260411A. All errors represent the 1$\sigma$ uncertainties.}
\label{tab:spec}
\begin{tabular}{cccccccccc}
\hline
\hline
\multirow{2}{*}{Data} & \multirow{2}{*}{Model} & $N_{\rm H}$ & $\Gamma_1$ & $\Gamma_2$ & $E_{\rm b}$ & $E_{\rm p}$ & log$A$ & stat/dof & BIC \\
& & ($\rm cm^{-2}$) & & & (keV) & (keV) & ($\rm{photons~cm^{-2}~s^{-1}~keV^{-1}}$) &  & \\
\hline
WXT & \textit{tbabs*pl} & $1.87\times10^{20}$ (fixed) & ${0.60}_{-0.66}^{+0.79}$ & - & - & - & ${-0.59}_{-0.19}^{+0.17}$ & 0.393/1 & 2.59 \\
GBM & \textit{cpl} & - & - & ${1.61}_{-0.12}^{+0.12}$ & - & ${568}_{-293}^{+514}$ & ${0.69}_{-0.21}^{+0.18}$ & 58.680/48 & 70.48\\
GBM & \textit{pl} & - & - & ${1.69}_{-0.08}^{+0.08}$ & - & - & ${0.79}_{-0.14}^{+0.15}$ & 57.556/49 & 65.42\\
WXT+GBM & \textit{tbabs*csbpl} & $1.87\times10^{20}$ (fixed) & ${0.11}_{-0.61}^{+0.40}$ & ${1.58}_{-0.16}^{+0.16}$ & ${7}_{-3}^{+6}$ & ${386}_{-178}^{+288}$ & ${-0.57}_{-0.19}^{+0.17}$ & 60.713/49 & 80.66 \\
WXT+GBM & \textit{tbabs*cpl} & ${5.25}_{-3.83}^{+5.67}\times10^{21}$ - & - & ${1.19}_{-0.20}^{+0.19}$ & - & ${252}_{-100}^{+296}$ & ${0.01}_{-0.26}^{+0.26}$ & 70.542/50 & 86.50 \\
\hline
\hline
\end{tabular}
\end{table*}

\end{document}